\documentclass[epj]{svjour}
\usepackage{graphics}
\begin{document}
\title{Determination of vector meson properties by matching resonance saturation to 
a constituent quark model}
\author{Stefan Leupold
}                     
\institute{Institut f\"ur Theoretische Physik, Universit\"at
Giessen, Germany}
\date{Received: date / Revised version: date}
%
\abstract{The properties of mesonic resonances can be calculated in terms of the
low-energy coefficients of chiral perturbation theory ($\chi$PT) by extending unitarized
$\chi$PT to higher energies. On the other hand these low-energy coefficients can be
calculated in two different models, namely (i) by assuming resonance saturation and 
(ii) within a constituent quark model. By matching the expressions of the two models 
combined with the results of unitarized $\chi$PT and the Weinberg sum rules
the properties of
vector and axial-vector mesons can be calculated in the combined large-$N_c$ and chiral
limit.
\PACS{
      {12.39.Fe}{Phenomenological quark models; chiral Lagrangians}   \and
      {14.40.Cs}{Properties of specific particles; other mesons with S=C=0, mass $<$ 2.5 GeV}
     } 
} 
\titlerunning{Vector meson properties}
\maketitle
\section{Introduction}
\label{intro}
\begin{figure}
\resizebox{0.45\textwidth}{!}{%
\includegraphics{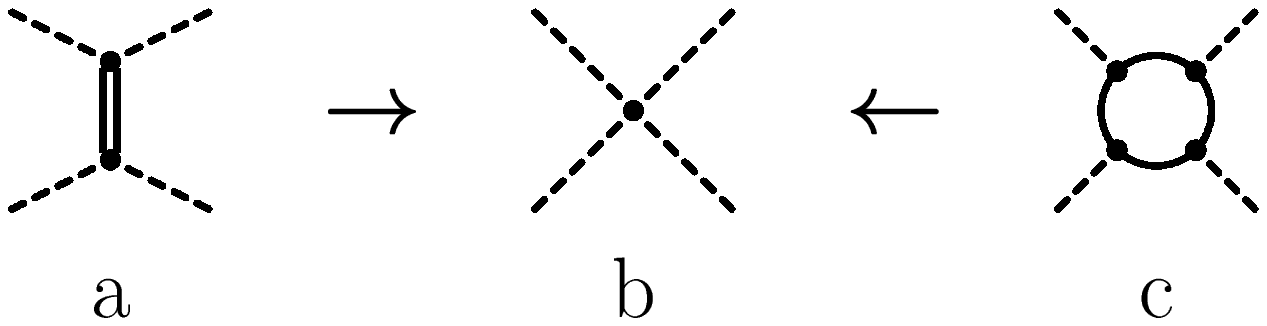}
}
  \caption{Schematic view of the resonance saturation model (a) and the constituent
quark model (c) and their respective low-energy reduction to $\chi$PT (b). The dashed
lines denote Goldstone bosons, the double line mesonic resonances and the full lines 
quarks.}
  \label{fig:fourpoint}
\end{figure}

What determines the properties of hadrons made from light quarks, chiral 
symmetry breaking ($\chi$SB) and/or confinement? It is nowadays common wisdom
that the mechanism of $\chi$SB causes large constituent quark masses of the order
of 300 -- 400 MeV. Hence even without confinement the creation of a quark-antiquark
pair is rather expensive. Therefore the role of confinement for the 
description of light hadrons is at least diminished by the appearance of $\chi$SB 
\cite{diak}. This suggests that 
the properties of light hadrons are {\em quantitatively} determined by the effect
of $\chi$SB. In such a scenario confinement enters only {\em qualitatively}
by excluding non-white states and quark-antiquark thresholds. It is well known
that such a picture works very well for pions (e.g.~\cite{NJL} and references therein). 
It is the purpose of the
present work to apply that picture to $\rho$- and $a_1$-mesons. One reason why one does
not need confinement to describe the properties of pions can be found in the fact that 
the mass of these quasi-Goldstone bosons is much below the (constituent!)
quark-antiquark threshold. This is of course different for other types of mesons.
At first glance it seems that this messes up the line of reasoning given above. 
The point however is 
that e.g.~$\rho$-mesons leave a trace also in the low-energy region much below their 
pole mass by mediating e.g.~pion-pion interactions \cite{donoghue}. 
Hence the key idea is that 
on the one hand side ($\chi$SB aspect) one can describe the low-energy region reliably 
by a (chiral!) quark model (without confinement) --- as this region is far away from the 
quark-antiquark production threshold. On the other hand side (confinement aspect)
the mesonic resonances are supposed to mediate the interactions in this low-energy
regime. By matching corresponding expressions it should be possible to determine
masses and coupling constants of mesonic resonances in terms of quark model expressions.
This procedure is depicted schematically in Fig.~\ref{fig:fourpoint}.
For simplicity I work in the following in the combined large-$N_c$ and chiral 
limit.\footnote{To be specific I take the large-$N_c$ limit first, i.e.~neglect the 
chiral log's which are suppressed by $1/N_c$.}

\section{Chiral perturbation theory and unitarization}
\label{sec:chipt}
At low energies QCD reduces to an 
effective theory where only the lightest mesons --- the pseudoscalar Goldstone
bosons --- appear which interact with each 
other and with external sources. $\chi$SB demands that 
the meson interaction vanishes with vanishing energy. Therefore a systematic 
expansion in terms of the derivatives of the meson fields is possible. These 
considerations lead to the effective lagrangian of chiral 
perturbation theory ($\chi$PT) \cite{gasleut}:
\begin{equation}
  \label{eq:chipt}
{\cal L}_{\chi \rm PT} = {\cal L}_1 + {\cal L}_2 + \mbox{higher order derivatives}
\end{equation}
with
\begin{eqnarray}
  \label{eq:chipt2}
{\cal L}_1 &=& {1\over 4} \,F_\pi^2 \,
{\rm tr} (\nabla_\mu U^\dagger \nabla^\mu U) + \ldots  \,, 
\\
{\cal L}_2 & = & L_1 \langle \nabla_\mu U^\dagger \nabla^\mu U \rangle^2 
+ L_2 \langle \nabla_\mu U^\dagger \nabla_\nu U \rangle 
      \langle \nabla^\mu U^\dagger \nabla^\nu U \rangle
\nonumber \\ && {}
+ L_3 \langle \nabla_\mu U^\dagger \nabla^\mu U \nabla_\nu U^\dagger \nabla^\nu U \rangle
\nonumber \\ && {}
-iL_9 \langle F_{\mu\nu}^{\rm R}\nabla^\mu U \nabla^\nu U^\dagger + 
               F_{\mu\nu}^{\rm L}\nabla^\mu U^\dagger \nabla^\nu U \rangle
  \label{eq:chipt4}
+ \ldots
\end{eqnarray}
where I have only displayed the terms which are relevant for later use. In $U$ the
pseudoscalar meson fields are encoded. 
$F_{\mu\nu}^{\rm R,L}$ denotes the field strength which
corresponds to (chirally covariant combinations of) external vector fields $v_\mu$
and axial-vector fields $a_\mu$. $F_\pi$ denotes the pion decay constant (in the chiral
limit). I refer to \cite{gasleut} for further details. The four-point meson 
interaction induced by (\ref{eq:chipt4}) is depicted schematically in 
Fig.~\ref{fig:fourpoint}b. 

As it stands the effective theory (\ref{eq:chipt}-\ref{eq:chipt4}) is valid at low 
energies only. 
Especially unitarity is not fulfilled. In \cite{pelaez} the inverse amplitude method
(IAM) is used to unitarize the effective theory and extend its applicability to the 
mesonic resonance region. The IAM is very well suited to recover a resonance from its
trace left at low energies \cite{oset} 
and therefore fits perfectly to the philosophy discussed
above. In \cite{pelaez} it is demonstrated that the IAM is able to reproduce the 
scattering data of pions, kaons and etas up to 1.2 GeV including several mesonic 
resonances. 
Here I am interested in the large-$N_c$ and chiral limit of the results presented
in \cite{pelaez}. It is easy to show that in this limit the mass of the $\rho$-meson
becomes \cite{jahnke}
\begin{equation}
  \label{eq:massrho}
M_V^2 = -{F_\pi^2 \over 4 L_3}  \,.
\end{equation}

\section{Chiral constituent quark model}
\label{sec:quark}
As already pointed out it should be reasonable to calculate the coefficients of the
effective theory (\ref{eq:chipt}-\ref{eq:chipt4})  from a chiral constituent quark model
as the low-energy region is (much) below the quark-antiquark production threshold. 
In the following I use the quark-Goldstone boson lagrangian (in euclidean space)
\begin{equation}
  \label{eq:lagrconst}
{\cal L}_{\rm quark} = 
\bar q \, (\gamma_\mu\partial_\mu - M U^{\gamma_5} + \gamma_\mu v_\mu + 
\gamma_\mu \gamma_5 a_\mu + \ldots)\,q
\end{equation}
This lagrangian can be motivated in several ways (e.g.~\cite{diakpet,espraf}). 
I would like to stress that it is also
the simplest model which one can write down which couples quarks to the Goldstone bosons
of $\chi$SB.
The latter are encoded in 
\begin{equation}
  \label{eq:defUg5}
  U^{\gamma_5} = {1-\gamma_5 \over 2} \, U + {1+\gamma_5 \over 2} \, U^\dagger  \,.
\end{equation}
$M$ denotes the mass of the constituent quark. 
The dots in (\ref{eq:lagrconst}) denote further couplings to external sources besides
the displayed ones for vector and axial-vector fields. 
By integrating out the quarks, expanding the obtained effective action in terms of meson
field derivatives and finally transforming the result to Minkowski space
one arrives at the $\chi$PT lagrangian (\ref{eq:chipt}) with predictions for the 
low-energy constants. This procedure is shown in Fig.~\ref{fig:fourpoint} as the
transition from Fig.~\ref{fig:fourpoint}c to Fig.~\ref{fig:fourpoint}b.
For the constants of interest one obtains \cite{espraf}
\begin{eqnarray}
L_1^{\rm quark} = {N_c \over 384 \pi^2} \,, && 
L_2^{\rm quark} = 2 L_1^{\rm quark} \,,
\nonumber \\
\label{eq:quarkpred}
L_3^{\rm quark} = - 4 L_1^{\rm quark}\,, &&
L_9^{\rm quark} =  8 L_1^{\rm quark} 
\,.
\end{eqnarray}
Note that the results are pure numbers, i.e.~do not depend on any
model-dependent quantities like e.g.~the UV-cutoff which regulates UV-singular integrals
coming from the non-renormalizable lagrangian (\ref{eq:lagrconst}). The reason is that
the loop depicted in Fig.~\ref{fig:fourpoint}c is actually UV-finite.

\section{Resonance saturation model}
\label{sec:ressat}
Assuming that (only) resonances mediate the low-energy interactions of (\ref{eq:chipt4})
one can write down a lagrangian which couples resonances to Goldstone 
bosons \cite{eckgas}. Here I concentrate on the $\rho$-meson and its interaction
lagrangian
\begin{equation}
  \label{eq:chirlagr}
{\cal L}_{\rm int} = {F_V \over 2 \sqrt{2}} \, {\rm tr}(V_{\mu\nu}f_+^{\mu\nu}) 
+ {i G_V \over \sqrt{2}} \, {\rm tr}(V_{\mu\nu}u^\mu u^\nu)
\end{equation}
where basically $u_\mu$ is obtained from $U$, i.e.~contains the pseudoscalar fields,
while $f^{\mu\nu}_{+}$ contains the external vector fields (see \cite{eckgas}
for details). $V_{\mu\nu}$ denotes the vector meson resonance
in the tensor representation. Note that one does {\em not} assume here that the vector 
meson couples with the same strength to the external vector 
field (photon) as it couples to the pseudoscalars (e.g.~pions). To phrase it differently,
universality of the $\rho$-meson coupling is {\em not an input} of the resonance
saturation model. As I will show below, however, one gains the universality as an 
{\em output} of my approach.
Integrating out the resonance fields one obtains (\ref{eq:chipt4}) with predictions
for the low-energy coefficients (schematically shown in Fig.~\ref{fig:fourpoint} 
by the transition from a to b).
For the ones governed solely by vector meson exchange one
gets \cite{eckgas}
\begin{equation}
  \label{eq:ressat}
L_2^{\rm res} = {G_V^2 \over 4 M_V^2} \,, \quad 
L_9^{\rm res} = {F_V G_V \over 2 M_V^2}  \,.
\end{equation}
Note that all the other low-energy constants are additionally influenced by the 
exchange of mesons with different quantum numbers.

\section{Results from matching}
\label{sec:match}
In the last sections I have basically collected results from the literature. The new 
aspect is now that the results from the approaches with hadronic degrees of freedom
(Secs.~\ref{sec:chipt}, \ref{sec:ressat}) are matched to the quark model calculations
of Sec.~\ref{sec:quark}. As already pointed out in the introduction the idea behind
that matching is that on the one hand side the chiral quark model is supposed to give
reliable results in the low-energy regime. On the other hand side confinement enforces
the formation of resonances (instead of the production of free quarks and antiquarks).
These resonances however are also visible at low energies, i.e.~determine the 
low-energy structure of the strong interaction. From the matching procedure one obtains
information about the resonances in terms of quark degrees of freedom. 

Even without the results of Sec.~\ref{sec:ressat}
one obtains from (\ref{eq:massrho}) and (\ref{eq:quarkpred}) for the $\rho$-meson mass:
\begin{equation}
  \label{eq:rhomassres}
M_V^2 = {24 \pi^2 F_\pi^2 \over N_c}  \,.
\end{equation}
Using the physical value for the pion decay constant $F_\pi$ $\approx$ $93\,$MeV
the $\rho$-meson mass turns out to be $M_V$ $\approx$ $826\,$MeV, already close to
the physical $\rho$-meson mass of $770\,$MeV. 
Note that this result was obtained in the chiral and large-$N_c$ limit. Hence pion
loops are absent in this framework, i.e.~I have determined the mass of a bare 
$\rho$-meson without its pion cloud. Typically the $\rho$-pion interaction reduces the 
bare $\rho$-mass by approximately 5 -- 10\% \cite{herrmann,klingl1}.

Using in addition (\ref{eq:ressat}) yields the coupling constants
\begin{equation}
  \label{eq:couplrhores}
F_V^2 = 2 F_\pi^2 \, \qquad  G_V^2 = {F_\pi^2 \over 2}  \,.
\end{equation}
In particular the relation $F_V = 2 G_V$ states the universality of
the vector meson coupling as can be most easily seen by inspecting (\ref{eq:chirlagr}).
This automatically implies that the KSFR relation is fulfilled \cite{weinb}.

The connection of $G_V$ to the usual $\rho\pi\pi$ coupling is provided by
\begin{equation}
  \label{eq:grhopipi}
g = {G_V M_V \over F_\pi^2}
\end{equation}
with $g$ defined via the lagrangian \cite{klingl1}
\begin{equation}
  \label{eq:standlagr}
{\cal L}_{\rm int} = {ig \over 4}\, {\rm tr}(V^\mu \, [\partial_\mu \Phi,\Phi])
- {g^2 \over 16}\, {\rm tr}([V^\mu,\Phi]^2)
\end{equation}
where $V^\mu$ is the vector meson resonance in the vector representation and
$\Phi$ is connected to $U$ via $U = \exp (i\Phi/F_\pi)$.
Relation (\ref{eq:grhopipi}) can be obtained by calculating the decay width
$\Gamma(\rho \to \pi\pi)$ in both approaches (\ref{eq:chirlagr}) 
and (\ref{eq:standlagr}). From (\ref{eq:rhomassres}), (\ref{eq:couplrhores}) one gets
\begin{equation}
  \label{eq:gres}
g = \sqrt{{3 \over N_c}} \, 2 \pi
\end{equation}
as compared to the experimental value of 6.05 
extracted from the decay width $\Gamma(\rho \to \pi\pi)$ \cite{klingl1}. 

Finally properties of the axial-vector meson $a_1$ can be deduced from the
Weinberg sum rules \cite{weinb,eckgas}:
\begin{equation}
  \label{eq:weinbersr}
F_V^2 = F_A^2 + F_\pi^2  \,, \qquad M_V^2 F_V^2 = M_A^2 F_A^2   \,.
\end{equation}
In combination with the previous results this yields the $a_1$-mass
\begin{equation}
  \label{eq:a1mass}
M_A = \sqrt{{3 \over N_c}} \, 4 \pi F_\pi \approx 1169\,\mbox{MeV}
\end{equation}
and the coupling of the $a_1$ to an external axial-vector current
\begin{equation}
  \label{eq:a1coupl}
F_A = F_\pi  \approx 93\,\mbox{MeV}
\end{equation}
to be compared to the experimental values $M_{a_1}$ = $1230\pm 40\,$MeV and
$F_{a_1}$ = $124\pm 27\,$MeV \cite{eckgas}.

\section{Summary and outlook}

I have presented a somewhat indirect way to determine the properties of vector and 
axial-vector mesons in terms of quark degrees of freedom. The success of the presented
approach suggests that it is indeed the phenomenon of $\chi$SB which
quantitatively determines the properties of the studied mesonic resonances. 
Confinement enters the framework only qualitatively by demanding that color-white
resonances are formed instead of quark-anti\-quark pairs. 

There are things which still have to be clarified: First, I have utilized two different
versions of resonance saturation. In the first one (Sec.~\ref{sec:chipt}) resonances
are created from a combination of the two lagrangians 
(\ref{eq:chipt2}) and (\ref{eq:chipt4}) (cf.~\cite{pelaez} for details) 
while in the second version 
(Sec.~\ref{sec:ressat}) the resonances only influence the fourth order lagrangian
(\ref{eq:chipt4}). The connection of these two versions has to be studied in more detail.
Second, there is a low-energy constant, namely $L_{10}$, which can be 
calculated both from the quark model and from the resonance saturation model using
vector and axial-vector mesons. It turns out that one needs more than one meson per
channel to achieve an agreement between the two calculations \cite{leupold02}. 

I expect that the presented framework can be extended from the vacuum case studied
here to the case of a medium with finite temperature and/or quark density. 
This should
provide interesting insight in the in-medium changes e.g.~of the $\rho$-meson mass,
its coupling to pions and photons and possible differences between longitudinal
and transverse $\rho$-mesons.

\end{document}